\documentclass[twocolumn,showpacs,preprintnumbers,amsmath,amssymb]{revtex4}

\usepackage{graphicx}
\usepackage{dcolumn}
\usepackage{bm}
\begin{document}
\preprint{APS/123-QED}

\title{Temperature dependent asymmetry of the nonlocal spin-injection
resistance:
evidence for spin non-conserving interface scattering}

\author{Samir Garzon$^1$\footnote{sgarzon@physics.sc.edu}, Igor \v{Z}uti\'{c}$^{1,2}$\footnote{zutic@dave.nrl.navy.mil}, and Richard A. Webb$^{1,3}$}

\affiliation{$^1$Department of Physics and Center for
Superconductivity Research, University of Maryland,
College Park, MD 20742\\
$^2$Center for Computational Materials Science, Naval Research
Laboratory, Washington D.C. 20375\\
$^3$Department of Physics and USC NanoCenter, University of South
Carolina, Columbia, SC 29208}


\begin{abstract}
We report nonlocal spin injection and detection experiments on
mesoscopic Co-Al$_2$O$_3$-Cu spin valves. We have observed a
temperature dependent asymmetry in the nonlocal resistance between
parallel and antiparallel configurations of the magnetic injector
and detector. This strongly supports the existence of a
nonequilibrium resistance that depends on the relative orientation
of the detector magnetization and the nonequilibrium magnetization
in the normal metal providing evidence for increasing interface
spin scattering with temperature.
\end{abstract}

\maketitle


Spin injection from a magnetic material is accompanied by a
nonequilibrium magnetization or spin accumulation near the
interfacial region. The interest to study spin accumulation, one
of the key elements in spintronic
applications~\cite{Zutic2004:RMP}, is not limited to novel
spin-based devices. Spin injection can also be used as a sensitive
spectroscopic tool to study fundamental properties such as the
pairing symmetry of unconventional
superconductors~\cite{Vasko1997:PRLetal,Ngai2004:Petal}, Skyrmion
excitations in the quantum Hall
regime~\cite{MacDonald1999:PRL,Chan1999:PRLetal}, and spin-charge
separation in non-Fermi liquids~\cite{Si1998:PRL,Balents2000:PRL}.
While spin accumulation in metals was first demonstrated by
Johnson and Silsbee~\cite{Johnson1985:PRL}, the basis for
understanding spin injection and, more generally, spin-polarized
transport, dates back to Mott~\cite{Mott1936:PRCa}. He noted that
the electrical current in ferromagnets could be expressed as the
sum of two independent and unequal parts for two different spin
projections implying that the current is spin-polarized. This
concept of a ``two-current model" together with the spin-dependent
scattering at the magnetic interfaces has been successfully used
to explain giant magnetoresistance (GMR) and tunnelling
magnetoresistance (TMR), key elements in applications such as
magnetic hard drives and nonvolatile magnetic random access memory
(MRAM)~\cite{Maekawa:2002a,Parkin2003:PIEEEetal}.

A conventional picture for spin injection across the
ferromagnet/nonmagnetic metal ($F/N$)
interfaces~\cite{Johnson1987:PRB,Rashba2002:EPJ,Takahashi2003:PRB}
is provided by noting that, in each metal, the spin-flip
scattering is typically much weaker than the momentum scattering.
This leads to a mean free path (MFP) $l$ which is much shorter
than the spin diffusion length (SDL) $\lambda$--the characteristic
scale for
the decay of spin accumulation at 
each side of the interface. Within the two-current model one can
then define local spin-resolved electrochemical potentials
$\mu_\sigma$, $\sigma=\uparrow,\downarrow$ for carriers with
majority and minority spin (with magnetic moment parallel and
antiparallel to the magnetization $\textbf{M}$ in a ferromagnet).
In the steady state an electrical current driven across a $F/N$
junction will lead to spin accumulation $\propto
(\mu_\uparrow-\mu_\downarrow)$ which is the balance between spins
added by the magnetization current and spins removed by spin
relaxation. In the absence of interfacial spin-flip scattering,
the spin-resolved current $I_\sigma$ is conserved across the
interface~\cite{Rashba2002:EPJ,Takahashi2003:PRB}, and the contact
resistance for each spin can be expressed as
\begin{equation}
R_\sigma=(\mu^F_\sigma-\mu^N_\sigma)/e I_\sigma,
\label{equ:rsigma}
\end{equation}
where the indices $F,N$ label the quantities in the corresponding
region at each side of the contact and $e$ is the proton charge.
$R_\uparrow \neq R_\downarrow$ can be inferred from the effect of
exchange splitting in the $F$ region, leading to spin-dependent
Fermi wave vectors, transmission coefficients, and density of
states. Spin accumulation in the $N$ region can act as a source of
spin electromotive force which produces a voltage $V \propto
(\mu_\uparrow-\mu_\downarrow)$ measurable by adding another
ferromagnet~\cite{Silsbee1980:BMR}.

\begin{figure}[b]
\scalebox{0.8}{\includegraphics[width=10.5cm,angle=0]{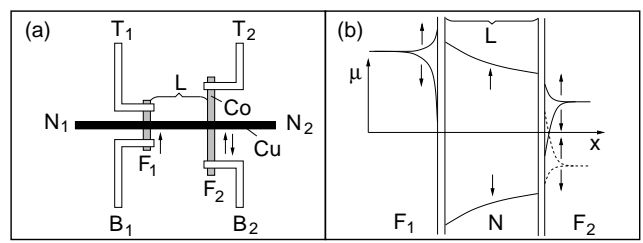}}
\caption{\label{fig:schematic} (a) Sample geometry. Shaded (black)
regions represent two Co ferromagnets (Cu line). $T_i$ and $B_i$
are nonmagnetic measurement leads. (b) $\mu_\sigma$ for spin up
and spin down electrons in injector (left), normal metal (center),
and detector (right) as a function of position. Detector
$\mu_\sigma$ for parallel (solid) and antiparallel (dashed)
magnetizations is shown.}
\end{figure}

Using the $F_1/N/F_2$ geometry depicted in
Fig.~\ref{fig:schematic}(a) in which $F_1$ represents the spin
injector and $F_2$ the spin detector, we performed nonlocal
measurements of spin injection
over a wide temperature range. 
Charge current is driven between the leads $T_1$ and $N_1$ while
the nonlocal voltage $V_{NL}$ is measured between the leads $T_2$
and $N_2$ which, in the absence of nonequilibrium spin, is an
equipotential region without a charge current flow such that
$V_{NL}=0$. As compared to local measurements (current driven
between $T_1$ and $T_2$ and voltage measured between $N_1$ and
$N_2$), the nonlocal measurement has been shown to simplify the
extraction of spurious effects (for example, anisotropic
magnetoresistance and the Hall effect) from those intrinsic to
spin injection~\cite{Johnson1985:PRL,Jedema2002:Naetal}. In
Fig.~\ref{fig:schematic}(b) we sketch a spatial profile of
$\mu_\sigma$ where the presence of interfacial scattering leads to
the discontinuity of the electrochemical potential
$\mu=(\mu_\uparrow+\mu_\downarrow)/2$ at each
contact~\cite{Rashba2002:EPJ,Takahashi2003:PRB,Valet1993:PRB}.

For $F_i/N_i$, $i=1,2$ tunnel contacts (with resistance much
larger than the characteristic values of the products $\rho_N
\lambda_N/A_N$ and $\rho_F \lambda_F/A_F$ with $\rho$ the bulk
resistivity and A the cross-sectional area) conventional analysis
gives a simple expression for the nonlocal
resistance~\cite{Johnson1987:PRB,Jedema2002:Naetal,Takahashi2003:PRB}
\begin{equation}
\label{equ:nonlocal} R_{NL} =\frac{V_{T_2-N_2}}{I_{T_1-N_1}} =\pm
\frac{\rho_N \lambda_N \exp(-L/\lambda_N)}{2 A_N} P_1 P_2,
\end{equation}
where the 
signs ``+" and ``-" refer to the parallel
($\uparrow \uparrow$)  and antiparallel ($\uparrow \downarrow$)
orientation of magnetizations in $F_1$ and $F_2$. $P_i$ is the
single F/N interface polarization of the current at contacts
$F_i/N_i$~\cite{pdef}, which can be expressed as the ratio of
spin-resolved contact resistances~\cite{Jonker2003:MRSetal}

\begin{equation}
\label{equ:defp}
P_i=(R_{i\uparrow}^{-1}-R_{i\downarrow}^{-1})/(R_{i\uparrow}^{-1}+R_{i\downarrow}^{-1}).
\end{equation}

Previous analysis of experimental data on spin injection in
metallic systems has only considered the spin-valve effect due to
the difference $R^{\uparrow \uparrow}_{NL}$-$R^{\uparrow
\downarrow}_{NL}$ between the nonlocal resistances  for $\uparrow
\uparrow$ and $\uparrow \downarrow$ orientation of magnetizations
in $F_1$ and
$F_2$~\cite{Johnson1985:PRL,Jedema2001:N,Jedema2002:Naetal,Johnson1988:PRBb}.
This approach was based on the understanding provided by either
Eq.~(\ref{equ:nonlocal})  or from a more general
expression~\cite{Takahashi2003:PRB}, not limited to the regime of
tunnel contacts, which shows that the symmetric combination
$R^{\uparrow \uparrow}_{NL}$+$R^{\uparrow \downarrow}_{NL}$
vanishes identically.

In contrast, our findings suggest that both the symmetric and the
antisymmetric combination of $R_{NL}$
\begin{equation}
R_{S,A}=(R^{\uparrow \uparrow}_{NL} \pm R^{\uparrow \downarrow}_{NL})/2,
\label{equ:sa}
\end{equation}
provide information about effects intrinsic to spin injection such
as interfacial spin scattering. Our measurements of $R_S$ show
that the usual assumption of $R_\sigma$ being equal for $\uparrow
\uparrow$ and $\uparrow \downarrow$ orientation does not hold.

Our samples are electron beam defined $F/N/F$ structures with
Co-Al$_2$O$_3$-Cu $F/N$ tunnel contacts as shown in
Fig.~\ref{fig:schematic}(a). $36$ nm of Co was thermally
evaporated to form the magnetic injector and detector. An Ar
ion-mill was used to clean the magnetic contacts before the $2$ nm
Al insulating level was thermally evaporated and oxidized for 4
minutes at pressures between $150$ mTorr and $515$ mTorr to obtain
different contact resistances. Finally $54$ nm of Cu was thermally
evaporated to complete the $F/N$ contacts. The width of the Cu
line was $100$ nm while the injector and detector were $105$ nm by
$2.4$ $\mu$m and $95$ nm by $4.5$ $\mu$m respectively to ensure
different coercivities allowing individual manipulation of their
magnetization. The devices were measured at $4.2$ K and between
$77$ K and room temperature~\cite{resist}. Typical values of the
SDL of Cu and Co are much larger than the MFP, so the macroscopic
diffusion equations are valid~\cite{Takahashi2003:PRB}. The
contact resistance of both magnetic contacts was measured at $4.2$
K and at room temperature, and found to be nearly temperature
independent and between $9$ $\Omega$ and $1$ k$\Omega$, depending
on the oxidation parameters, so the approximation $R_i>>\rho
\lambda/A$ is valid for the $F$ and $N$ regions.

\begin{figure}
\begin{center}
\scalebox{0.8}{\includegraphics[width=10.5cm,angle=0]{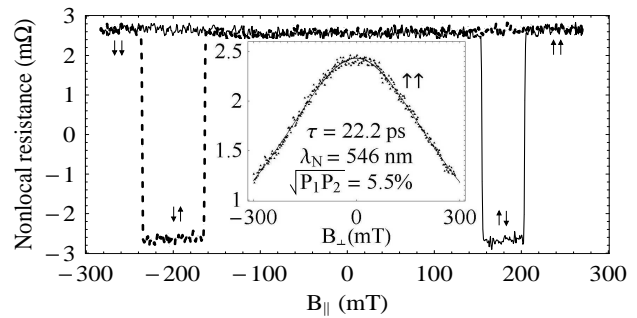}}
\caption{\label{fig:steps} $R_{NL}$ as a function of parallel
field at $4.2$ K for an injector-detector separation of 430 nm.
The solid (dashed) trace is for increasing (decreasing) magnetic
field. The arrows describe the direction of $\textbf{M}$ of
injector and detector. Inset: Hanle effect for parallel injector
and detector demonstrating precession of injected spins.}
\end{center}
\end{figure}

AC currents between $10$ $\mu$A and $50$ $\mu$A were injected from
$T_1$ to $N_1$, see Fig.~\ref{fig:schematic}(a). Both the nonlocal
voltage between $T_2$ and $N_2$, and the voltage at a $100$
k$\Omega$ bias resistor were measured using lock-in amplifiers at
$11$ Hz. First $R_{NL}$ was measured at $4.2$ K as the magnetic
field parallel to the injector and detector ($B_\parallel$) was
cycled between $-300$ mT and $300$ mT at $0.0003$ Hz. We observed
the characteristic switching of the nonlocal resistance whenever
the magnetic contacts became parallel or antiparallel [Fig.~\ref{fig:steps}, Eq.~(\ref{equ:nonlocal})]. 
The magnitude of $R_{NL}$ for $\uparrow\uparrow$ and
$\uparrow\downarrow$ magnetic contacts is almost the same,
$|R_{NL}^{\uparrow\uparrow}|\approx |R_{NL}^{\uparrow\downarrow}|$
so at low temperatures $R_S$ vanishes.



Hanle effect measurements~\cite{Zutic2004:RMP} were performed to
determine the degree of spin polarization and the spin diffusion
length $\lambda_N$ of electrons in Cu. The magnetic contacts were
aligned by applying $B_\parallel$. After $B_\parallel$ was turned
off, an out-of-plane magnetic field $B_\perp$ was applied to
induce spin precession as the electrons diffused from F$_1$ to
F$_2$ with a time distribution $P(t)=1/\sqrt{4\pi D t} \exp(-L^2/4
D t)$. Increasing $|B_\perp|$ also increases the precession angle
of the electron spin making its projection onto $\textbf{M}$ of
F$_2$ smaller, hence decreasing $R_{NL}^{\uparrow\uparrow}$.
Diffusive motion of the electrons leads to a broad distribution of
transport times and precession angles so a weighted average of the
right hand side of Eq.~(\ref{equ:nonlocal}) must be used. The
Hanle resistance, equivalent to the solution of Bloch-Torrey's
equations~\cite{Zutic2004:RMP,Johnson1988:PRBb,Jedema2002:Naetal},
is

\begin{equation}
\label{equ:hanle} R_{H}=\pm \frac{P_1 P_2 D
\rho_N}{A_N}\int_0^\infty P(t) \cos (\omega t) e^{-D
t/\lambda_N^2} dt,
\end{equation}

\noindent where $\omega=g \mu_B B_\perp/\hbar$ is the Larmor
frequency. The measured Hanle effect signal is weakly asymmetric
(inset of Fig. \ref{fig:steps}) possibly because the two magnetic
contacts were not perfectly parallel, or because the applied field
had some nonzero in-plane component. Similar data was obtained for
the case of antiparallel injector and detector. By fitting the
Hanle effect data to Eq.~(\ref{equ:hanle}) we found that the spin
diffusion length was $546$ nm and the product of the spin
polarizations of the two magnetic contacts was $\sqrt{P_1
P_2}=5.5\%$. The value of the SDL is comparable to the values
measured using GMR~\cite{Yang1994:PRLetal} ($450$ nm) and
transparent $F/N/F$ spin valves~\cite{Jedema2001:N} ($1000$ nm).
The values of the spin polarization are also in agreement with
those measured using $F/N/F$ spin valves~\cite{Jedema2002:Naetal}.

\begin{figure}[t]
\scalebox{0.8}{\includegraphics[width=10.5cm,angle=0]{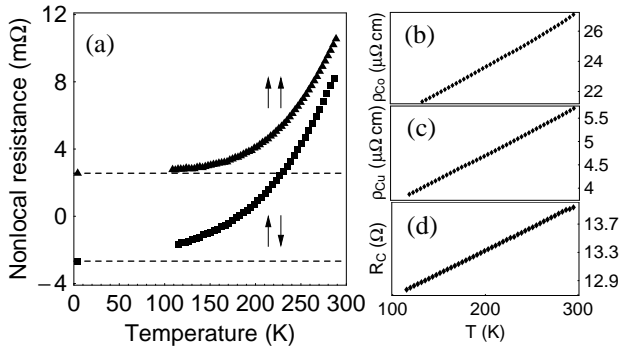}}
\caption{\label{fig:tempdep}(a) Temperature dependence of the
nonlocal resistance for parallel (triangles) and antiparallel
(squares) injector and detector. Dashed lines represent $4.2$ K
values. (b),(c) Temperature dependence of the resistivity of Co
and Cu. (d) Typical temperature dependence of the contact
resistance.}
\end{figure}

\begin{figure}[t]
\scalebox{0.8}{\includegraphics[width=10cm,angle=0]{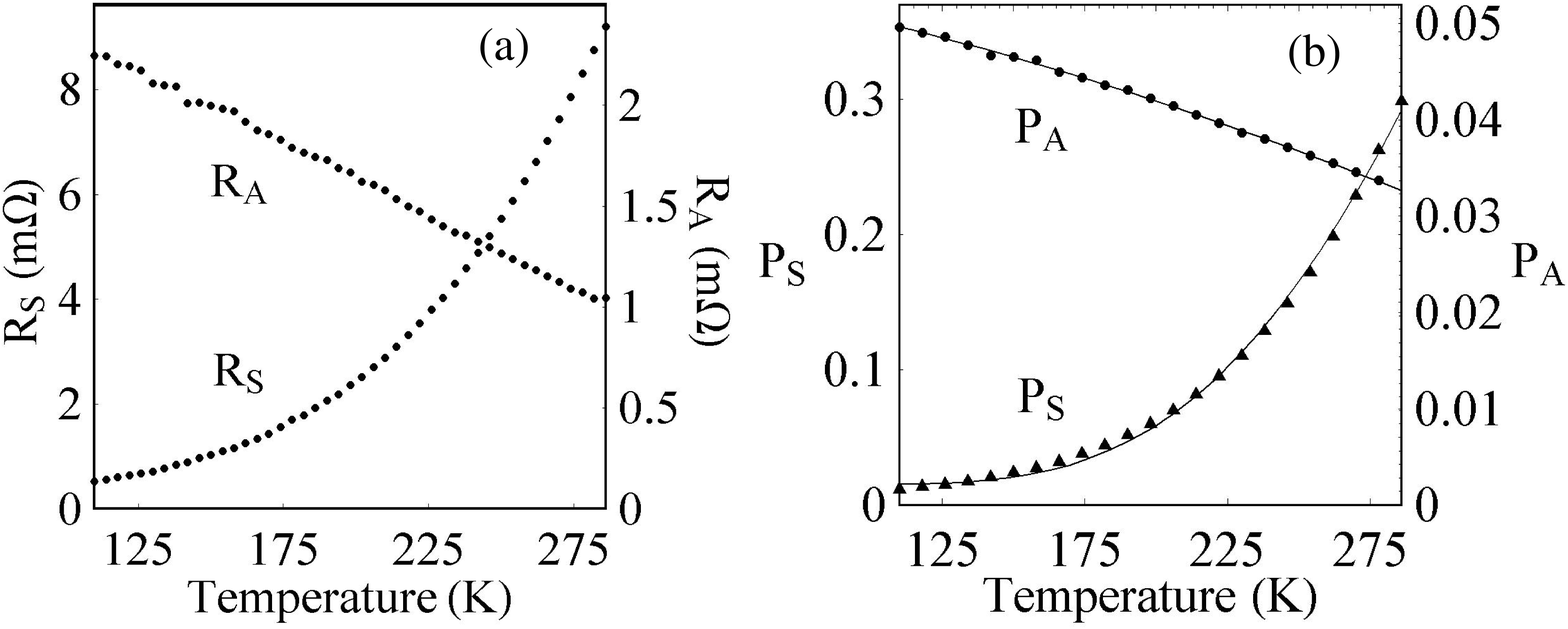}}
\caption{\label{fig:symasym} Temperature dependence of the
symmetric and antisymmetric components of (a) $R_{NL}$, and (b)
$P_2$ together with the fits from the proposed model.}
\end{figure}

We measured $R_{NL}(T)$ for 100K$<$T$<$300K and observed a
temperature dependent asymmetry between
$R_{NL}^{\uparrow\uparrow}$ and $R_{NL}^{\uparrow\downarrow}$ as
shown in Fig.~\ref{fig:tempdep}(a). Our data shows that $R_S=0$
only at low temperatures but at room temperature $R_S$ is three
times larger than the low temperature value of $R_{NL}$. The
temperature dependence of $R_S$ and $R_A$ is shown in
Fig.~\ref{fig:symasym}(a) to illustrate that $R_A$ decreases
linearly in this temperature range while the increase in $R_S$ is
nonlinear. This provides some evidence of the possibly different
origin of $R_A$, which is magnetization-dependent, and $R_S$,
which is independent of the relative magnetization of injector and
detector. We see the
expected~\cite{Jedema2001:N,Jedema2002:Naetal} trend of decreasing
$R_A$ as the temperature increases due to the increase in spin
scattering in Cu and the reduction of the magnetization of Co due
to magnons. For comparison we also measured $\rho(T)$ of Co and
Cu, which increased linearly for 100K$<$T$<$300K [Figs.~\ref{fig:tempdep}(b),(c)], 
while $R_{NL}$ increased nonlinearly in that range
[Fig.~\ref{fig:tempdep}(a)]. Since $P_1$ should not change when
$\textbf{M}$ of the detector is reversed, the origin of $R_S$ can
be traced to the difference in $P_2$ between $\uparrow\uparrow$
and $\uparrow\downarrow$ configurations [see
Eq.~(\ref{equ:nonlocal})]. Hence the quantities
$P_{S,A}=(P_2^{\uparrow\uparrow}\pm P_2^{\uparrow\downarrow})/2$
contain the relevant information of the spin transport across the
N/F$_2$ interface. From previous studies~\cite{Shang1998:PRBetal}
$P_A$ should have the same temperature dependence as the
magnetization, $P(T)=P_0(1-\eta T^{3/2})$. Since the momentum
relaxation time $\tau_p\propto\tau$ (the spin relaxation time)
where $\tau_p/\tau=a^p$~\cite{Zutic2004:RMP}, we can use
Fig.~\ref{fig:tempdep}(c) to evaluate $\lambda_N(T)$. This,
together with $P(T)$ and the measured value of the spin
polarization from Hanle effect ($P_0=5.5\%$), can be used to fit
$R_A$ with Eqs.~(\ref{equ:nonlocal}) and~(\ref{equ:sa}).
Figure~\ref{fig:symasym}(b) shows the obtained $P_A$ together with
a fit to $P(T)$ with $\eta=8.4\times 10^{-5}$K$^{-3/2}$ and
$a^p=6.6\times 10^{-4}$. As expected from the interface effects
$\eta$ is larger than the bulk value~\cite{Shang1998:PRBetal} and
$a^p$ agrees very well with the previously measured
value~\cite{Jedema2001:N}. With the same two fitting parameters it
is possible to extract the temperature dependence of $P_S$ from
the measured $R_S$, as shown in Fig.~\ref{fig:symasym}(b), which
can be fit well by a Fermi function that describes thermal
activation with a characteristic temperature of 1227 K. The
measured offset $R_S$ can be modelled by assuming that the
spin-dependent contact resistances depend on the relative
alignment of the nonequilibrium magnetization in the normal metal
and the magnetization in the ferromagnetic contact, so there are
four independent contact resistances that describe the transport
across each interface. This was shown to be possible in
semiconductors due to nonlinear effects but the argument is not
valid in metals~\cite{Zutic2002:PRL}. The most likely explanation
for $R_S\neq 0$ is the existence of increasing spin-dependent
scattering at the detector interface with temperature. In contrast
with TMR, where the same charge current travels through injector
and detector, in the nonlocal geometry the absence of charge
current at the detector justifies treating $F_1$ and $F_2$
differently. We confirmed this behavior by solving the diffusion
equations including spin scattering at the $F_2/N$
interface~\cite{barnas}. Equation~(\ref{equ:nonlocal}) still holds
if an effective spin polarization is redefined as $P_2 \rightarrow
\tilde{P_2}$, where~\cite{pm}

\begin{equation}
\label{equ:defp2}
\tilde{P_2}=\frac{(R_{2\uparrow}^{-1}-R_{2\downarrow}^{-1})
\pm(R_{2\uparrow}'^{-1}-R_{2\downarrow}'^{-1})}
{(R_{2\uparrow}^{-1}+R_{2\downarrow}^{-1})
+(R_{2\uparrow}'^{-1}+R_{2\downarrow}'^{-1})}=P_{A}\pm P_{S}.
\end{equation}

The primed quantities, absent in Eq.~(\ref{equ:defp}), represent
the spin-scattering contact resistances~\cite{Rashba2002:EPJ}
which allow the existence of non-vanishing $R_S$ and
$P_S$~\cite{rdef}. Even though $P_A$ includes spin conserving and
spin-scattering resistances, Fig.~\ref{fig:symasym}(b) shows
clearly that it has the expected $1-\eta T^{3/2}$ dependence. This
is clarified by Fig.~\ref{fig:tempdep}(d) which shows that the
measured contact resistance, proportional to the inverse of the
denominator of Eq.~(\ref{equ:defp2}), is weakly T dependent for
100K$<$T$<$300K. Therefore the strong T dependence of $P_S$ comes
from the difference in the interfacial spin scattering of the two
spin channels at the detector
($R_{2\uparrow}'^{-1}-R_{2\downarrow}'^{-1}$), which can be
expected at a magnetic interface. While a detailed quantitative
confirmation requires a fully microscopic picture, the increase of
up to 30\% in the relative scattering of spin $\uparrow$ and
$\downarrow$ at $F_2$ for 100K$<$T$<$300K can be expected given
the large spin polarization ($>$50\%) of surface states of
Co~\cite{Okuno2002:PRL}.

Figure~\ref{fig:tempdep} rules out spurious voltages offsets that
can appear from the electrostatics of the cross geometry of the
injector~\cite{Johnson1986:T}, since the value of that voltage
would depend linearly on the contact resistance and the
resistivity of both the Cu line and the Co injector, which are
linear in temperature, as opposed to the measured offset which
increases nonlinearly. Samples in which one or both of the Co
contacts had been replaced by Cu showed $R_S \approx 0$, ruling
out artifacts coming from the setup. Increasing the separation $L$
between $F_1$ and $F_2$ caused the value of $R_S$ to decrease and
approach zero for large values of $L$ ($\approx 2 \mu$m), implying
that $R_S$ is produced by the interplay of $F_1$ and $F_2$.

We verified these measurements by repeating them at DC, studied
the response at higher frequencies, and replaced the lock-in
amplifiers by a spectrum analyzer. We also checked the linearity
of the signal with respect to the bias current. Leakage currents
through the input impedance of the lock-in were negligible, and
capacitive effects were ruled out by the DC measurements. All the
results were consistent with those shown above. Heating, coupled
with the difference in Seebeck coefficients of the two metals, can
generate spurious voltages. However, these voltages should only
appear at twice the fundamental frequency since the power goes as
I$^2$. We studied this possibility by looking at the second
harmonic response using both lock-in measurements and a spectrum
analyzer, and were able to see a signal $\propto$ I$^2$.
Furthermore, we replaced Co by Cr whose relative Seebeck
coefficient with respect to Cu is opposite in sign, and were able
to see a change in the sign of the I$^2$ dependence, confirming
the existence of thermal voltages at the second harmonic and not
at the fundamental frequency.

In conclusion, temperature dependent measurements of the nonlocal
resistance in $F_1/N/F_2$ junctions reveal a previously overlooked
asymmetry in the contact resistance for different relative
orientations of the magnetizations in ferromagnets $F_1$ and
$F_2$. We performed systematic control experiments to rule out a
spurious origin of this asymmetry due to the electrostatic field
distribution at imperfect contacts, Joule heating and
magneto-thermal effects. Our analysis suggests that a conventional
interpretation of spin injection which assumes spin-conserving
interfaces needs to be generalized~\cite{rashba}. We developed a
phenomenological model to show that the observed data is
consistent with temperature dependent spin-flip interfacial
scattering. We believe that additional theoretical and
experimental work will be required before a complete understanding
of this new spin-flip effect is obtained. 

This work was supported by NSF, NSA Laboratory for Physical
Science, and NRC (I.\v{Z}.). We thank M. Johnson and E. I. Rashba
for useful discussions.

\bibliography{references_offset}

\begin{thebibliography}{32}
\bibitem[{pde()}]{pdef}
\bibinfo{note}{In contrast with TMR, $P_2$ should not be interpreted as the
  spin current polarization at the detector since in the nonlocal geometry no
  charge current reaches $F_2$ and therefore the spin current polarization at
  $F_2$ would be infinite. By definition $P_2$ is just the spin current
  polarization at $F_2/N$ when $F_2$ is used as an injector.}
\bibitem[{res()}]{resist}
\bibinfo{note}{The resistivity $\rho$ of the Cu line was $3.44$ $ \mu \Omega$cm
  at $4.2$ K and $5.76$ $ \mu \Omega$cm at room temperature. Control samples
  for resistance determination of Co were fabricated under the same conditions
  as the spin-valve structures. Typical values for $\rho$ of Co were $18.7$ $
  \mu \Omega$cm at $4.2$ K and $27.2$ $ \mu \Omega$cm at room temperature. The
  corresponding mean free path and diffusion coefficient of Cu were $25.7$ nm
  and $1.34\times 10^{-2}$ m$^2$/s at $4.2$ K and $15.3$ nm and $0.80\times
  10^{-2}$ m$^2$/s at room temperature.}
\bibitem[{ras()}]{rashba}
\bibinfo{note}{The efficiency and the temperature dependence of spin injection
  may likely be controlled by the spin-flip rate at the interface (a transport
  quantity) rather than, as it is usually assumed, by an estimated Curie
  temperature of an injector (a thermodynamic quantity).}
\end{thebibliography}

\end{document}